\documentclass{article}

\usepackage[dblblindworkshop, final]{neurips_2025}
\workshoptitle{AI for Music}

\usepackage[utf8]{inputenc} 
\usepackage[T1]{fontenc}    
\usepackage{hyperref}       
\usepackage{url}            
\usepackage{booktabs}       
\usepackage{amsfonts}       
\usepackage{nicefrac}       
\usepackage{microtype}      
\usepackage{xcolor}         
\usepackage{graphicx}

\title{Chord-conditioned Melody and Bass Generation} 

\author{%
  Alexandra C Salem\\
  CUNY Graduate Center\\
  New York, NY 10016 \\
  \texttt{asalem1@gradcenter.cuny.edu} \\
  \And
  Mohammad Shokri \\
  CUNY Graduate Center\\
  New York, NY 10016 \\
  \texttt{mshokri@gradcenter.cuny.edu} \\
  \AND
  Johanna Devaney \\
  Brooklyn College / CUNY Graduate Center\\
  Brooklyn, NY 11210 /  New York, NY 10016\\
  \texttt{johanna.devaney@brooklyn.cuny.edu} \\
}

\begin{document}

\maketitle

\begin{abstract}
We evaluate five Transformer-based strategies for chord-conditioned melody and bass generation using a set of music theory–motivated metrics capturing pitch content, pitch interval size, and chord tone usage. The evaluated models include (1) no chord conditioning, (2) independent line chord-conditioned generation, (3) bass-first chord-conditioned generation, (4) melody-first chord-conditioned generation, and (5) chord-conditioned co-generation. We show that chord-conditioning improves the replication of stylistic pitch content and chord tone usage characteristics, particularly for the bass-first model. 

\end{abstract}

\section{Introduction}

Musicians typically receive their first exposure to music through listening, followed by playing. Those trained in the Western art music tradition are subsequently taught skills for analyzing the harmonic and structural aspects of music and creating model compositions in the music theory classroom using formal pedagogies. Musicians use these analytic skills in composition exercises, where they emulate a particular musical style. These exercises are often built on an existing chord progression.

In this paper, we compare the performance of chord-conditioned music generation strategies on a style composition task. We focus on melody and bass generation, as a standard approach in composition exercises is to outline the bass and melody before "filling-in" the inner voices in between them. Also, the melodic and bass lines are more characteristic than the inner voices. Melodic lines tend to move more by step (whole tones or semitones), while bass lines move by step and by leaps (intervals larger than a whole tone). We use the stylistically homogeneous TAVERN \citep{devaney15tavern} theme and variation dataset for training and evaluation music generation in the high Classical style. We train five different Transformer-based generative models: (1) independent melody and bass generation without chord information, (2) independent melody and bass generation with a chord progression supplied, (3) sonority-by-sonority bass note and then melody note generation with a chord progression supplied, (4) sonority-by-sonority melody note and then bass note generation with a chord progression supplied, and (5) co-generation of bass and melody lines with a chord progression supplied. The five models are compared to held-out TAVERN data using objective metrics that approximate the criteria used for assessing human-created model compositions (pitch choice, intervallic size, and chord tone usage). 



There is some existing work on chord-conditioned Transformer models for symbolic music generation. These include monophonic (single voice) generation 
\citep{choi_chord_2021,li_chord_2023}; 
and polyphonic (bass and melody line) generation \citep{shu_chord_2024} from chords in a popular music style. 
In a Classical music style, and thus most similar to our work, \cite{liu2022symphonygenerationpermutationinvariant} developed a Transformer model for multi-instrument multi-track symphonic scores. However, they evaluated their work with only subjective human ratings, rather than music theory-motivated objective metrics, nor did they specifically evaluate how well the generated scores aligned with the supplied chords. 
In this work, our contribution is a Transformer based chord-conditioned model for generating both bass and melody in Classical music, alongside objective music theory-based evaluation of the generated music.

\section{Methodology}

\subsection{Data}

The Theme and Variation Encodings with Roman Numerals (TAVERN) corpus \citep{devaney15tavern} consists of 27 complete sets of theme and variations for piano composed by Mozart and Beethoven in high Classical style, divided into musical phrases. Within each theme and variation set, the harmony remains relatively constant across the theme and variations, while the theme's melody is embellished in the variations through changes in rhythm, tempo, texture, key, and mode across different musical ``textures''. Each TAVERN phrase score is represented in **kern format \citep{huron2002music} with four ``spines'' (i.e., columns): function, harmonic, bass voice, and melody voice. The spines are aligned across time, allowing mapping of which notes in each voice occurred during each annotated chord. We split the data into train (25 sets consisting 1,702 phrases) and test (2 sets---one Beethoven, one Mozart---consisting of 164 phrases) so that we could evaluate the models on entirely unseen sets. 

We used score reductions of the **kern files to reduce the task complexity. We filtered each voice to include one quarter note per annotated chord: for the bass voice we filtered to the lowest pitch per chord, and for the melody voice we filtered to the highest pitch per chord. If a note was not present for the chord, a quarter rest was included. This process resulted in melody and bass line skeletons that captured the overall shape of the lines. Additionally, we transposed all scores from their original keys to C major or C minor. Next, we converted the scores to MIDI, one of the most commonly used machine-readable score formats. During model preparation, the scores were tokenized into REMI format \citep{huang2020pop}, which represents a MIDI score as a sequence of pitch, velocity, and duration tokens. The corresponding chords for each score were represented as a Harte-style pop chords \citep{Harte2010} (e.g., ``C:min G:maj G:maj C:min'') with chord inversions removed. These simplified chords were mapped to an integer between 1 and 24 and then tokenized.

\subsection{Model Architectures and Experiments}

\begin{figure}
  \centering
  \includegraphics[width=\textwidth]{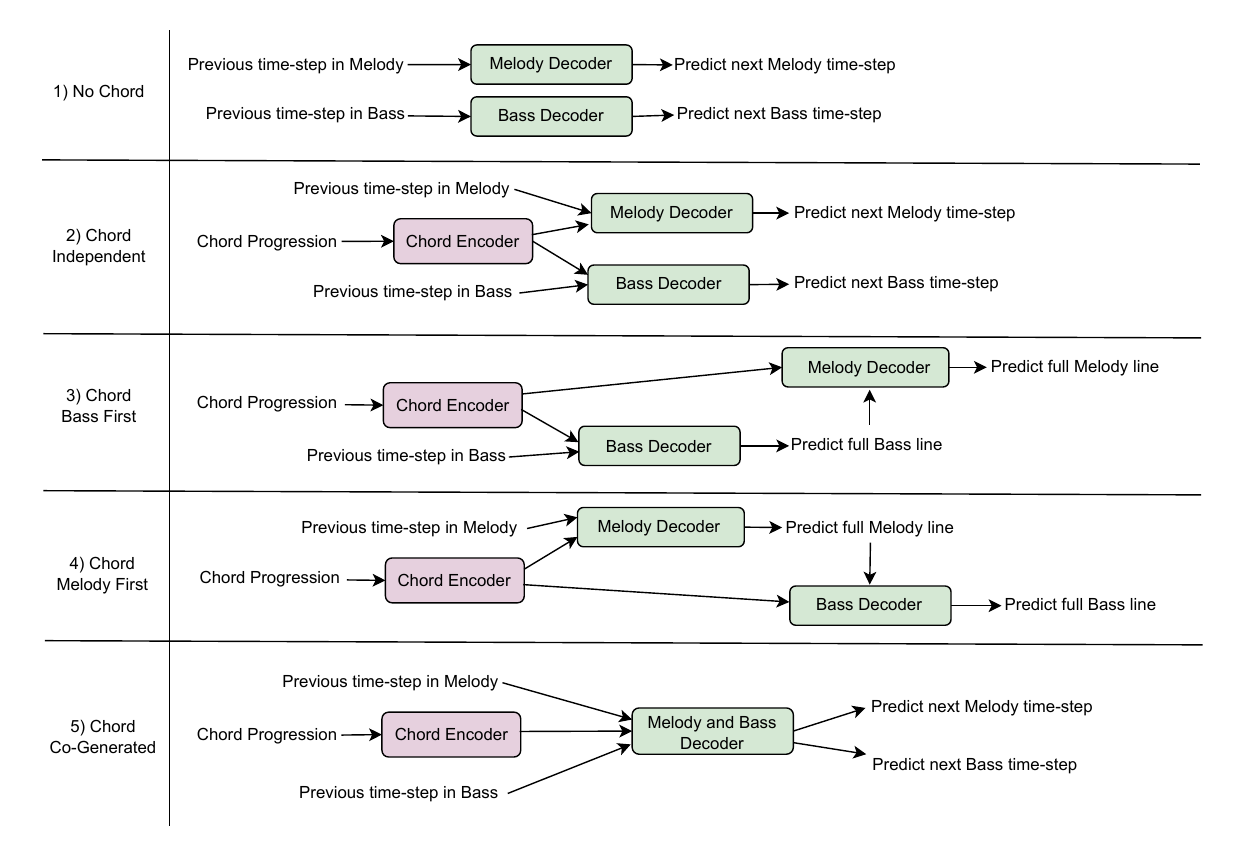}
  \caption{Visualization of the model variants.}
  \label{fig1}
\end{figure}

We implemented five model variants (see Figure \ref{fig1}) in order to compare different generation strategies of the melody and bass line skeletons. 
All models use variants of the Transformer architecture~\citep{transformer}, with either a decoder-only or encoder-decoder structure. 
A maximum length of 128 was used for chord and score inputs, with padding or truncation as necessary. All chord encoders used one encoder layer and two attention heads per layer, and all REMI decoders used six decoder layers and eight heads per layer. We used cross-entropy loss, and a learning rate of 1e-4, with the Adam optimizer and the lambda learning rate scheduler with 1000 warmup steps. Models were trained for 400 epochs with a batch size of 8. During inference, top-p decoding was used, with $p = 0.9$. The code, further details, and a subset of generated samples are in our GitHub repository\footnote{\url{https://github.com/alexandrasalem/symbolic-music-generation}}. 

The first variant (No Chord) performs melody and bass generation without any chord information. This model uses a decoder-only architecture, in which the model predicts next REMI tokens given previously seen ones. The second (Chord Independent) is wholly independent chord-conditioned generation of the bass and melody lines, where each voice is generated separately from one another. In this architecture, a chord encoder first produces a representation of the chord progression. Then, a decoder (either bass or melody) predicts a REMI token given the chord progression representation and the previously generated REMI tokens. These two models serve as baselines for the other variants, where chord-conditioned melody and bass lines are generated in a non-independent manner.
In the third (Chord Bass-1st), the output of the chord encoder gets passed to a bass decoder, which predicts the entire bass sequence of REMI tokens, similarly to Chord Independent. Then the entire bass line prediction is passed alongside the chord encoder output to a melody decoder, which then predicts the melody tokens. In the fourth (Chord Melody-1st), the model is the same, but the melody notes are generated first, and then the bass notes.
In the fifth (Chord Co-Gen), the bass and melody are co-generated, conditioned on the chords. In this architecture, the chord encoder output gets passed to a single decoder, which then predicts both the bass and melody REMI tokens concurrently.

\subsection{Evaluation}
We selected evaluation metrics for our experiment that approximate how human-generated model compositions are evaluated. We divide our metrics into three categories. The first captures information about the pitch content of the melody and bass lines. This category consists of the number pitch classes used in the melody or bass line, divided by the number of notes present (PCs Used), which gives us information about pitch class diversity;  the pitch class histogram entropy across each melody or bass line \citep{huang2020pop} (PC Entropy), which gives information about the distribution of the pitch classes; the number of unique pitches used in the melody or bass line \citep{zhang2020learning} (Unique Pitches), which gives information about the diversity of pitches (where pitch classes in different octaves are counted as different); the pitch range in the melody or bass line \citep{yang2020evaluation} (Pitch Range), which gives information about how spread out the pitches are in frequency; the unique pitch class ratio between melody and bass (Unique PC Ratio), which gives information about whether the same or different chord tones are being generated for a given time-step in the melody and bass; and Pitch Consonance Score (PCS), which captures consonance between bass and melody \citep{yeh2021automatic}. The second category captures information about pitch interval size, which we use to determine how closely the models are replicating the step-heavy characteristic of melodies versus the more leap-heavy characteristic of bass lines. Specifically, we calculate the average pitch interval size across each melody and bass line \citep{yang2020evaluation} (Pitch Interval). And the third captures information about the use of chord tones versus non-chord tones in the melody and bass lines by calculating the chord tone to non-chord tone ratio across each whole melody or bass line \citep{yeh2021automatic} (CT Ratio). 
All of the metrics are calculated separately on the generated melody and bass lines, except for Unique PC Ratio and PCS, which are calculated between the melody and bass lines.  We use the Wilcoxon signed-rank test to evaluate whether the differences between the ground truth test data and the generated data across model variants are statistically significant.

\section{Results}
Table \ref{tab1} shows the metrics for our model-generated melody and bass lines, averaged across the TAVERN held-out test set. The metrics for the  TAVERN test set itself, our ground truth, are also reported. Bold typeset indicates a metric distribution is \textit{not} significantly different from its corresponding ground truth distribution, by the Wilcoxon signed-rank test with $\alpha = 0.008$, accounting for multiple comparisons. For our purposes we consider this to be an indication that the generated musical lines are stylistically consistent with the ground truth. 
Chord Bass-1st had the most robust performance across all three metric categories, achieving statistically indistinguishable results from the ground truth for Pitch Interval, CT Ratio, and the majority of pitch content metrics.
Chord Mel-1st and Chord Ind. were close behind, but both performed worse on the CT Ratio metric than Chord Bass-1st and Mel-1st also performed worse on Pitch Interval. Chord Co-Gen failed to replicate many ground truth pitch content qualities, as well as CT ratio, although it did perform well on Pitch Interval.
Finally, No Chord was the most different from the ground truth, with only PC Entropy (Mel.), Unique PC Ratio, and Pitch Interval (Mel.) being significantly indistinguishable from the ground truth.




\begin{table}
  \caption{Experimental results for the five model variants (\textit{N}=164). Bolding indicates that the result for a metric is not statistically significantly different from the ground truth.}
    \label{results-table}
  \centering
\begin{tabular}{llllllll}
\toprule
Metric & Voice & No  & Chord & Chord & Chord & Chord & GT\\
 &  & Chord &  \emph{Ind.} &  \emph{Mel-1st} & \emph{Bass-1st}&  \emph{Co-Gen} &  \\
\midrule
PC Entropy & Mel. & \textbf{1.456}         & 1.627           & 1.574           & 1.665           & 1.773           & 1.506 \\
& Bass & 1.210                  & \textbf{1.540}  & \textbf{1.523}  & \textbf{1.423}  & \textbf{1.547}  & 1.474\\
PCs Used & Mel. & 0.519                  & \textbf{0.501}  & 0.521           & \textbf{0.493}  & 0.528           & 0.482\\
& Bass & 0.433                  & \textbf{0.491}  & \textbf{0.501}  & \textbf{0.464}  & \textbf{0.476}  & 0.471\\
Unique Pitches & Mel. & 5.805                  & \textbf{7.494}  & \textbf{6.683}  & 9.421           & 8.841          & 7.152\\
& Bass & 5.280                  & 7.695           & \textbf{7.207}  & \textbf{6.811}  & 7.927          & 7.335 \\
Pitch Range & Mel. & 12.634                 & \textbf{15.323} & \textbf{13.939} & 19.945          & 16.884         & 17.390 \\
& Bass & 14.177                 & \textbf{17.976} & \textbf{17.646} & \textbf{18.049} & \textbf{17.573} & 15.211\\
Unique PC Ratio & & \textbf{0.853}         & \textbf{0.836}  & \textbf{0.839}  & 0.840           & 0.864          & 0.820 \\
PCS & & 0.248 & \textbf{0.360} & 0.261 & 0.333 & 0.319 & 0.402\\
\midrule
Pitch Interval & Mel. & \textbf{3.356}         & \textbf{3.296}  & 3.298           & \textbf{4.216}  & \textbf{3.274}  & 4.467\\
& Bass & 4.001                  & \textbf{4.104}  & \textbf{4.423}  & \textbf{4.128}  & 3.654          & 3.731 \\
\midrule
CT Ratio & Mel. & NA & 0.728           & 0.714           & \textbf{0.716}  & 0.744           & 0.662 \\
& Bass & NA & \textbf{0.676}  & 0.717           & \textbf{0.683}  & 0.721    & 0.611 \\     
\bottomrule
\end{tabular}
\label{tab1}
\end{table}

\section{Conclusions and Future Work}
This work presented a set of chord-conditioned Transformer models for melody and bass generation in simplified high Classical style. 
All models clearly benefited from the information in the chords, as the No Chord model saw the lowest performance.
Our best-performing model across all metric groups was Chord Bass-1st, which learned to first predict the bass line and then the melody. This aligns with how students are typically instructed to do this task. Somewhat surprisingly, Chord Co-Gen did not perform as well as the Chord Bass-1st model, suggesting that bass-melody generation ordering is more beneficial in replicating this style of music than jointly estimating bass and melody notes for each time step.

In future work, we will test additional model variations. Reflecting upon Chord Co-Gen's performance, we may see improvement if each melody or bass prediction step is conditioned on the preceding time step's bass note, melody note, and chord instead of only the preceding bass \textit{or} melody note and entire chord progression, as in the current implementation. This change would combine the improvements we saw in the Chord Bass-1st model, with the benefits of a single decoder structure.
Importantly, we will also conduct a subjective evaluation with experts to complement the presented objective metrics. 
This work presents a first step in a pedagogy-inspired musical score generation approach that could benefit downstream tasks such as chord-controlled training data generation for analysis tasks and creation of educational exercises.

\begin{ack}
This work is supported by the National Science Foundation (NSF) award 2228910. 
\end{ack}

\bibliographystyle{plainnat}
\bibliography{refs}

\end{document}